\documentclass[twocolumn,showpacs,floatfix,prl]{revtex4}

\usepackage{graphicx}
\usepackage{dcolumn}
\usepackage{bm}
\usepackage{amsmath}
\usepackage{amssymb}

\usepackage{ulem}

\renewcommand{\xout}[1]{}

\usepackage{color}
\definecolor{PSRed}{rgb}{0.7,0.1,0.1} 
\definecolor{TURed}{rgb}{0.9,0.1,0.1}

\newcommand{\etal}{\mbox{\textit{et~al.}} } 

\newcommand{\ket}[1]{\left| #1 \right\rangle}
\newcommand{\bra}[1]{\left\langle #1 \right|}

\begin{document}

\graphicspath{{.}}

\title{Color-charge separation in trapped SU(3) fermionic atoms}

\author{Tobias Ulbricht$^{1}$}
\author{Rafael A. Molina$^{2}$}
\author{Ronny Thomale$^{3}$}
\author{Peter Schmitteckert$^{4}$}


\affiliation{
  $^1$Institut f\"ur Theorie der Kondensierten Materie, Karlsruhe Institute of Technology, 76128 Karlsruhe, Germany \\
  $^2$Instituto de Estructura de la Materia - CSIC, Serrano 123, 28006 Madrid, Spain\\
  $^3$Department of Physics, Princeton University, Princeton, NJ 08544, USA\\
  $^4$Institut f\"ur Nanotechnologie, Karlsruhe Institute of Technology, 76344 Eggenstein-Leopoldshafen, Germany }

\begin{abstract}
  Cold fermionic atoms with three different hyperfine states with
  SU(3) symmetry confined in one-dimensional optical lattices show
  color-charge separation, generalizing the conventional spin charge
  separation for interacting SU(2) fermions in one dimension. %
  Through time-dependent DMRG simulations, we explore the features of
  this phenomenon for a generalized SU(3) Hubbard Hamiltonian. %
  In our numerical simulations of finite size systems, we observe
  different velocities of the charge and color degrees of freedom when
  a Gaussian wave packet or a charge (color) density response to a
  local perturbation is
  evolved. 
  The differences between attractive and repulsive interactions are
  explored and we note that neither a small anisotropy of the
  interaction, breaking the SU(3) symmetry, nor the filling impedes
  the basic observation of these effects. %
\end{abstract}


\maketitle

One of the most intriguing effects of strong correlations in
low-dimensional systems is the separation of charge and spin degrees
of freedom. %
As a generic phenomenon, a quantum particle carrying spin and charge
converts in separate spin (spinon) and charge (holon) excitations with
generally different velocities. On a microscopic bare level, there are
rare examples where this process can be exactly studied in terms of
explicit ($n$ species) spinon and holon wave functions, as is the case
for the Kuramoto-Yokoyama
model~\cite{PhysRevLett.67.1338,PhysRevB.65.195112,PhysRevB.74.024423,PhysRevB.75.024405}.
However, on a low-energy effective field theory level, spin-charge
separation (SCS) explicitly manifests itself in all generic
one-dimensional interacting systems belonging to the Luttinger liquid
universality class, where spinons and holons are described by
independent collective density-type excitations
\cite{Giamarchi_BOOK2004}. %
In spite of several attempts, the {\it direct} experimental observation of
SCS has proved elusive
\cite{Bockrath_Cobden_Lu_Rinzler_Smalley_Balents_McEuen_N1999,Segovia_Purdie_Hengsberger_Baer_N1999,Losio_Altmann_Kirakosian_Lin_Petrovykh_Himpsel_PRL2001,Lorenz_Hofmann_Gruninger_Freimuth_Uhrig_Dumm_Dressel_N2002}.
So far, the best experimental evidence is provided by tunneling
between quantum wires where interference effects relate to the
existence of excitations with different velocities
\cite{Auslaender_Steinberg_Yacoby_Tserkovnyak_Halperin_Baldwin_Pfeiffer_West_S2005}.

Since recently, one possible avenue for the observation and
exploration of SCS are ultracold Bose or Fermi gases confined in
optical lattices that have become an important instrument for
investigating the physics of strong correlations. %
The great advantage of these systems is that the interaction parameters and
dimensionality can be tuned with very high precision and control by
means of an atomic Feshbach resonance or by changing the depth of the
wells in an optical lattice
\cite{Lewenstein_Sanpera_Ahufinger_Damski_Sen_Sen_AIP2007}. %
One of the first achievements in this subject is the experimental
observation of a superfluid to Mott insulator transition in a
three-dimensional optical lattice with bosonic $^{87}$Rb atoms
\cite{Greiner_Mandel_Esslinger_Hansch_Bloch_N2002}. %
Interesting experimental results have also been obtained for
fermions, like the study of the molecule formation through the
Feshbach resonance
\cite{Stoferle_Moritz_Gunter_Kohl_Esslinger_PRL2006}, the experimental
observation of fermionic
superfluidity\cite{Chin_Miller_Liu_Stan_Setiawan_Sanner_Xu_Ketterle_N2006},
and the observation of a Mott insulator in an optical lattice
\cite{Jordens_Strohmaier_Gunter_Moritz_Esslinger_N2008}. %
Several theoretical works have addressed the possibility of observing
SCS in cold-fermionic gases
\cite{Kollath_Schollwock_Zwerger_PRL2005,Schmitteckert_Schneider_HPC2006,Kollath_Schollwock_NJP2006,Schmitteckert_HPC2007,Ulbricht_Schmitteckert_EPL2009,Ulbricht_Schmitteckert_EPL2010}
and in cold-bosonic gases
\cite{Kleine_Kollath_McCulloch_Giamarchi_Schollwock_PRA2008} with two
internal degrees of freedom. %
Thanks to the special properties of cold atomic systems these
theoretical proposals could address previously unexplored features of
SCS, like the problem of the breaking of SU(2)
symmetry and SCS at high energies. %
In particular, higher spin fermions can be directly studied with cold atoms in more
than two hyperfine states.\xout{ without any substantial increase of complication.} %
This kind of systems could give rise to new phases in optical lattices
due to the emergence of triplets and quartets (three or four fermion
bound states), and other phenomena
~\cite{Honerkamp_Hofstetter_PRL2004,Wu_PRL2005,Lecheminant_Boulat_Azaria_PRL2005,Kamei_Miyake_JPSJ2005,Rapp_Zarand_Honerkamp_Hofstetter_PRL2007,gorshkov,PhysRevB.80.180420}. %
At least two alkali atoms $^6$Li and $^{40}$K seem to be possible
candidates for the experimental realization of an SU(3) fermionic
lattice with attractive
interactions~\cite{Rapp_Zarand_Honerkamp_Hofstetter_PRL2007}. %
In the case of $^6$Li the scattering lengths for the three possible
channels of the three lowest lying hyperfine levels ($\ket{F,m} =
\left|1/2, 1/2\right>$, $\left|1/2,-1/2\right>$, and
$\left|3/2,-3/2\right>$) at large magnetic fields become similar for
the three of them $a_s \approx -2500 a_0$
\cite{Bartenstein_Altmeyer_Riedl_Geursen_Jochim_Chin_Denschlag_Grimm_Simoni_Tiesinga_Williams_Julienne_PRL2005}.
Moreover, the realization of a stable and balanced three-component
Fermi gas has been recently reported to potentially accomplish both an
attractive and repulsive regime with approximate SU(3)
symmetry~\cite{Ottenstein_Lompe_Kohnen_Wenz_Jochim_PRL2008,PhysRevLett.102.165302}. %
The scattering lengths of the different channels for the three lowest
hyperfine states of $^{40}$K near the Feshbach resonance was also
measured and the possibility of trapping them optically was
demonstrated~\cite{Regal_Thesis2005}. %

It is the purpose of this  work to use time-dependent
Density Matrix Renormalization Group (td-DMRG) 
\cite{White_PRL1992,Schmitteckert_PRB2004,Daley_Kollath_Schollwoeck_Vidal_JSMTE2004,White_Feiguin_PRL2004,RevModPhys.77.259}
simulations to study the
phenomenology of CCS in lattice systems with three
different kinds of fermions, where the color notation is inherited from the quark description of high energy SU(3) theories of quantum chromodynamics.
As one of the significative quantities extractable from td-DMRG, the
different color and charge velocities are taken out from
time-dependent simulations, and compared to a low energy bosonization
approach. %


The low energy physics of cold fermionic atoms with three different
hyperfine states trapped in an optical lattice can be described by an
SU(3) version of the Hubbard Hamiltonian,
\begin{eqnarray}
  H&=&-t\sum_{\left\langle ij\right\rangle, \alpha} \left(f_{i\alpha}^{\dagger}f^{}_{j\alpha }+{\mathrm{h.c.}}\right)+ \nonumber
  \sum_{i,\alpha\neq\beta} \frac{U_{\alpha\beta}}{2} n_{i\alpha}n_{i\beta}.
\label{eq:Hamiltonian}
\end{eqnarray}
The sums $\alpha$ and $\beta$ extend over the three colors red(r),
green(g), and blue(b) corresponding to three different hyperfine
states. %
The operators $f_{i\alpha}^{\dagger}$ and $f^{}_{i\alpha }$ are the
creation and destruction operators of an atom on site $i$ with color
$\alpha$. %
We consider different values of the on-site interaction between the
different color pairs $U_{\alpha\beta}$ to be able to include SU(3)
symmetry breaking terms. %
The site label $i$ goes from $0$ to $L-1$, with $L$ being the total
number of lattice sites corresponding to the wells forming the optical
lattice. %
For cold atoms, there is an additional harmonic confinement term that
arises due to the Gaussian profile of the laser beams. %
If this trap potential is weak, we can assume to sit in the trap
center where the confinement can be considered constant. %
Larger potentials could be taken into account, \xout{ by a basis change,}
e.g. see \cite{Ulbricht_Schmitteckert_EPL2010}. 
In the subsequent calculations\xout{,} we will ignore the confinement. %
The hopping term can be controlled by varying the depth of the wells
through the laser intensity. %
The optical lattice potential that each of the hyperfine states is affected by 
is very similar and the hopping rates can be considered to be equal
for the three different colors. %
In the rest of the paper, all energies will be expressed in units of
$t=1$. %
The interaction strength in each channel is proportional to the
corresponding s-wave scattering length $U_{\alpha \beta}=4 \pi \hbar^2
a_{\alpha \beta}/m$. %
The condition for the atoms to stay in the lowest band is that the
s-wave scattering length must be smaller than the typical size of the
wave function of an atom in one of the lattice wells
\cite{Jaksch_Bruder_Cirac_Gardiner_Zoller_PRL1998}. %
In turn, this is easily fulfilled in the experiments, so we will neglect the
population of higher bands. %
Spin-flipping rates are usually smaller than the escape rate of atoms
from the optical lattices and can be neglected as well. %


In the elementary linear bosonization approach, valid in the
weak-coupling limit, the low-energy effective theory of the model can
be expressed in terms of the collective fluctuations of the densities
of the three spin species plus charge density. %
Introducing the three bosonic fields $\phi_\alpha(x)$, the density
operators for each color can be written as
\begin{equation}
  \rho_{i,\alpha} \approx \bar{\rho}+\frac{1}{\sqrt{\pi}} \partial_x \phi_\alpha(x)-\frac{1}{\pi} \sin{\left[ 2k_Fx+
      \sqrt{4\pi}\phi_\alpha(x)\right]},
\label{eq:density_fluctuations}
\end{equation}
where $k_F$ is the Fermi wave-vector and $x$ corresponds to the
lattice site. %
We can express the bosonized Hamiltonian in terms of the total density
described by a bosonic field
\begin{equation}
\Phi_{\text{ch}}=\frac{1}{\sqrt{3}}\sum_\alpha \phi_\alpha,
\end{equation}
and the relative density described by two bosonic fields
\begin{equation}
\Phi_{3}=\frac{1}{\sqrt{2}} (\phi_r-\phi_b)
\end{equation}
\begin{equation}
\Phi_{8} = \frac{1}{\sqrt{6}} (\phi_r+\phi_b-2 \phi_g).
\end{equation}
The subindices were chosen to correspond to the SU(3) group
Casimir operators $J_{3}$ and $J_{8}$. %
For a SU(3) symmetric Hamiltonian, the model can be separated into two
parts, charge and color, $H=H_{\text{ch}}+H_{\text{col}}$,
\begin{equation}
  H_{\text{ch}}=\frac{v_{\text{ch}}}{2}\left[\frac{1}{K}(\partial_x \Phi_{\text{ch}})^2+K(\partial_x \Theta_{\text{ch}})^2\right],
\end{equation}
where $v_{\text{ch}}$ is the density velocity $v_{\text{ch}}=v_F/K$,
$K=(1+2U/\pi v_F)^{-1/2}$ is the Luttinger liquid parameter, and
$\Theta_{\text{ch}}$ is the conjugate field of the bosonic field
$\Phi_{\text{ch}}$, and
\begin{align}
  H_\text{col} & =\sum_{\mu=3,8} \left[ \frac{v_F}{2}
    \left((\partial_x \Phi_{\mu})^2+(\partial_x \Theta_{\mu})^2
    \right)
    + \frac{U}{2\pi} (\partial_x \Phi_{\mu})^2 \right]  \nonumber \\
  & + \frac{U}{2 \pi^2} \cos{\sqrt{2 \pi} \Phi_3} \cos{\sqrt{6 \pi}
    \Phi_8} - \frac{U}{2 \pi^2} \cos{\sqrt{8 \pi} \Phi_3}.
\label{Hc}
\end{align}
Similarly, the color velocity can be approximated as
$v_{\text{col}}=v_F\sqrt{1 - U/(\pi v_{F})}$. %
However, due to the non-linear cosine terms in (\ref{Hc}), we do not
expect a linear relation between distance and time for color
excitations to hold for long times. %
If SU(3) symmetry is not strictly observed there appear mixing terms
coupling density and color degrees of freedom. %
For example, if $U_{rg}=U_{rb}=U_1$ and $U_{gb}=U_2$ the mixing term
can be written as $H_{\text{mix}}=\sqrt{2}(U_1-U_2)/{3 \pi} \cdot
\partial_x \Phi_{3} \partial_x \Phi_{8}$
\cite{Assaraf_Azaria_Caffarel_Lecheminant_PRB1999}. %
A renormalization group analysis of this model can elucidate the
low-energy properties of our system
\cite{Assaraf_Azaria_Caffarel_Lecheminant_PRB1999,PhysRevA.80.041604}.
The most important difference with respect to the SU(2) case is that
when $U>0$ but weak, umklapp processes present for commensurate
fillings do not open a gap in the charge sector. %
We expect a phase transition between the MI and the LL at a finite
value of $U$. %
In fact, with Monte-Carlo calculations, Assaraf \etal estimated the
critical interaction scale at $U_{\text{cr}}=2.2$
\cite{Assaraf_Azaria_Caffarel_Lecheminant_PRB1999}, while recent DMRG
calculations suggest the critical point to be much closer to zero
\cite{Buchta_Legeza_Szirmai_Solyom_PRB07}. %
The cosine terms for the color Hamiltonian are irrelevant for
repulsive, but relevant for attractive interaction and
responsible of a gap opening in the color sector in the attractive
case. %

Let us now elaborate on how the CCS appears in real-time simulations. We
study the time evolution of the Hamiltonian (\ref{eq:Hamiltonian})
with the td-DMRG algorithm
\cite{Schmitteckert_PRB2004,Ulbricht_Schmitteckert_EPL2009}. %
The number of states needed to keep sufficient accuracy during the
time evolution was more than $7000$.
Such huge demands limited the system size we could simulate to $L=18$
for periodic (PBC) and $L=48$ for hard-wall boundary conditions
(HWBC), which in total is settled in the range of present state
of the art limits. %
For the small systems, the finite size effects are large. Comparable
simulations on SU(2) systems provided detailed knowledge about the finite
size effects, so as that they are relatively
independent on the interaction parameter. %
Comparison with exact results for $U=0$ of the charge and color
velocities provide us with approximations to the Luttinger liquid
parameter.

\begin{figure}
  \includegraphics[width=\linewidth]{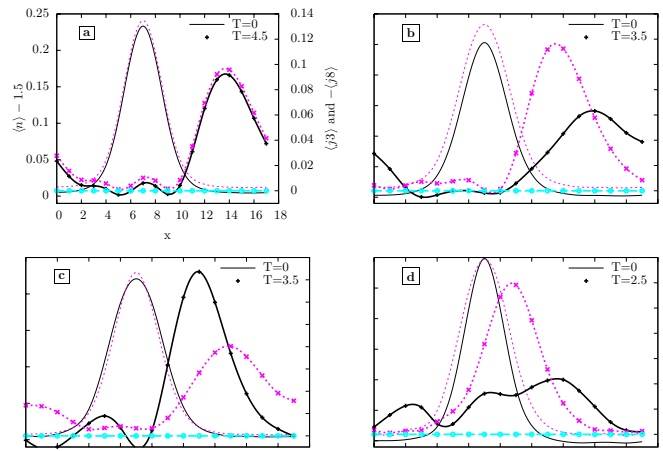}
  \caption{(color online) Snapshots of the time evolution of an
    additional fermion for filling $\nu=0.5$ and $U=0$ (a), $U=+1$
    (b), $U=-1$ (c) and $U=+5$ (d). %
    The initial state is a Gaussian wave packet with average momentum
    $k=0.6\pi$. %
    In (b,c,d) the color density of quantum number $j^{8}$ (magenta)
    separates from the charge density (black), while $j^{3}$ (cyan) is
    constant by construction. %
    Lines serve as guides to the eye. %
  }
  \label{fig:snapshots}
\end{figure}

We show snapshots of the time evolution for different interaction
strengths in Fig.~\ref{fig:snapshots} for the system with $L=18$ and
PBC. %
We calculate the ground state $| \Psi_{0} \rangle$ of the SU(3)
Hamiltonian with $N=27$ fermions and then put an extra (green) fermion
with an initial wave packet in our system that travels to the right
with a finite momentum centered around $k=0.6\pi$
\begin{equation}
\label{eq:wp}
\ket{\Psi^{+1}(t=0)}  = \sqrt{\frac{\pi}{2\sigma^2}} \sum_x \text{e}^{( i k x)} \text{e}^{-\frac{(x-x_0)^2}{2\sigma^2}} f^{\dagger}_x \ket{\Psi_0},
\end{equation}
where the width of the wave packet is chosen to be $\sigma=1.5$. %
Choosing $N=27$ leads to an average density of $\overline{\langle n
  \rangle}=1.5$ and an incommensurate filling of $\nu=0.5$, the
commensurate fillings being $\nu=1/3,2/3$. %
Time is always measured in inverse units of the hopping rate. %
The panels in the figure show the particle density $\langle n \rangle
= \bra{\Psi^{+1}(t)} \sum_{\alpha}
f_{\alpha}^{\dagger}f_{\alpha}\ket{\Psi^{+1}(t)}$ and the
corresponding densities of the color quantum numbers $j^{3}$ and
$j^{8}$ relative to the uniform ground state density at initial and a
finite time. %
We observe a slight dispersion effect due to \xout{to the finite size of our
system and} the finite width of the wave-packet. Moreover, the velocity
of the excitation is not exactly the velocity of a plane wave with
momentum $k$. %
However, as relative velocity quantities are concerned, without
interaction the charge density and color velocity are exactly the same
(upper left panel). %
In Fig. \ref{fig:snapshots}, for the case $U=1$, we start to see the
separation of density and color degrees of freedom. %
The initial excitation separates and the color and charge density
evolve with different velocities. %
The decay of the charge density excitation is more rapid for strong
repulsive interaction as it is seen in the case $U=5$, clearly
indicating the opening of the charge gap. In similar ways, a gap in
the color sector should open for attractive interaction, and
spoil the color density evolution in that regime. %
However, in our example with $U=-1$, we can still observe rather clean
and stable CCS, since, for our finite system, the arising gap in the
color sector is too small to reasonably detect an enhanced color
excitation decay.  %
To extract color and charge velocities, we applied two different
extraction methods. First, the velocity can be obtained dividing the
number of sites traveled by the maximum of the density by the time. %
It is more accurate to fit a Gaussian on the density for every time
step. %
There is a short transient time at the beginning, followed by a
plateau of constant velocity until the packet hits the boundary. %
We extracted the velocity of the packet when it passes the median
between the initial position and the boundary and took a Gaussian
averaged value around this position. %
As a second measure, for a definite plateau of constant velocity, we
used a $L=48$ sites system with HWBCs and
replaced the incident electron by an initial, small potential
perturbation. %
The time development of this method differs mainly by the implicit
stimulation of excited states around both Fermi points leading to
symmetric peaks running in both directions and increasing the
transient time at the start. %
\begin{figure}
  \includegraphics[width=\linewidth]{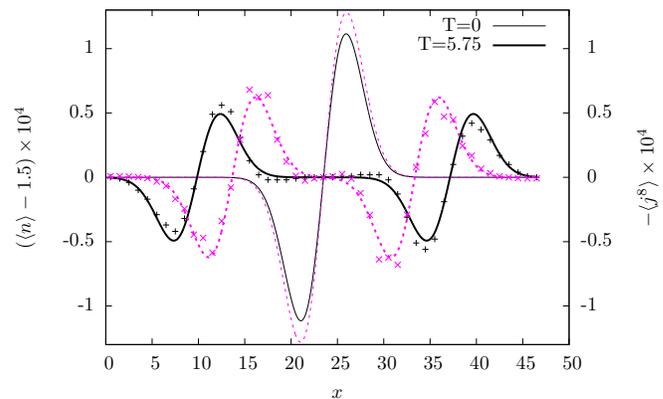}
  \caption{(color online) Snapshots of a perturbation at time $T=0$
    (thin lines) and a finite time step (thick lines and
    data points). %
    The dashed (solid) lines are fits to the color density of $j^8$
    (charge density $n$) with the corresponding data points.%
  }
  \label{fig:snapshot_potential}
\end{figure}
Fig. \ref{fig:snapshot_potential} shows snapshots of the initial and a
finite time step for $U=1.5$, where the perturbation was taken to be
the derivative of a Gaussian and this function was fitted on the two
propagating wave packets. %
We find that both methods provide similar results, while the latter
proved to be a considerable improvement in precision and is put
forward by us as a general preferable method to extract velocities
from td-DMRG.

\begin{figure}
  \includegraphics[width=\linewidth]{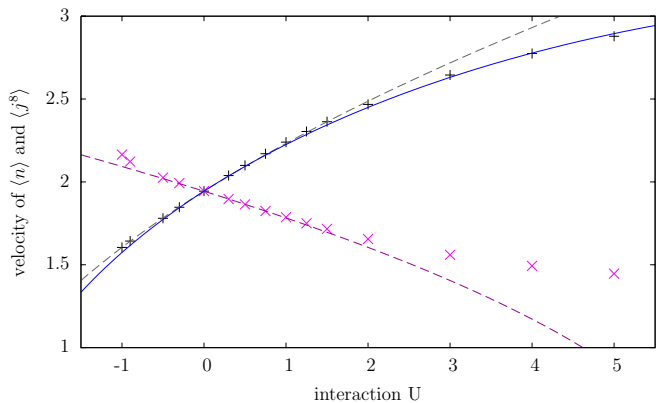}
  \caption{(color online) Velocity $v(U)$ of an initial Gaussian
    perturbation. %
    Results of our td-DMRG simulations for both charge density ($+$)
    and color density ($\times$). %
    The dashed lines represent bosonization results, scaled to match
    the DMRG velocity at $U=0$. %
    The solid line is a fit on a next-to linear dependency in $U$. %
  }
  \label{fig:vel}
\end{figure}

Finally, the extracted velocities are shown as a function of the
uniform interaction in Fig. \ref{fig:vel}. %
On top of the data, we show the expected relation for
$v_{\text{col}}(U), v_{\text{ch}}(U)$ from bosonization, scaled to
match the free fermion velocity. %
This {\it a priori} renormalization covers all direct effects that
lead to deviations of the Fermi velocity due to finite width or the
central momentum $k_{0}$ of the excitation and the finite size
dispersion. %
Even for $|U|$ up to $1$, the agreement between the bosonization
estimate and the numerical data is found to be excellent (dashed lines
in Fig. \ref{fig:vel}). %
Beyond, the limits of the low energy expansion are apparently reached.
However, upon extending the velocity expansion to higher order in $U$,
e.g.  $v_{\text{ch}} = v_{F}\sqrt{1 + 2U/(\pi v_{F}) + a U^{2}}$, the
fit already covers the complete computed range in $U$ (solid line in
Fig.  \ref{fig:vel}) and the parameter $a=-0.017$ has the same order
and sign as the fit by Assaraf \etal
\cite{Assaraf_Azaria_Caffarel_Lecheminant_PRB1999} who chose a
different filling in their Monte-Carlo simulations. %

A possible candidate for realizing the trionic
phase ($U < 0$) is $^6$Li for which the magnetic field dependence of the three
scattering lengths has been measured
\cite{Bartenstein_Altmeyer_Riedl_Geursen_Jochim_Chin_Denschlag_Grimm_Simoni_Tiesinga_Williams_Julienne_PRL2005}. %
The attractive $U$ interaction for magnetic fields around 1000 G can
be estimated as $U_{rg}=U_0$, $U_{rb}=1.23U_0$, and $U_{gb}=1.06U_0$,
due to the ratios in the s-wave scattering rate in each channel
\cite{Bartenstein_Altmeyer_Riedl_Geursen_Jochim_Chin_Denschlag_Grimm_Simoni_Tiesinga_Williams_Julienne_PRL2005}. %
It is important to consider whether these anisotropies infringe on the
validity of the effects observed above.  In bosonization theory,
anisotropy introduces new terms that couple the charge and color
degrees of freedom so we would expect that high anisotropies totally
destroy charge-color separation
\cite{PhysRevA.80.041604}. %
However, small and experimentally accessible anisotropies turn out to
be not decisively important. %
We have checked that the separation effect of color and charge
densities is still visible in the simulations without much change. %
Even for the special cases of commensurate fillings where a stronger sensitivity on the color anisotropy may have been suspected, the qualitative
behavior persists. %


In summary, we have performed td-DMRG simulations of cold
fermionic atoms with three hyperfine states trapped in an optical
lattice. %
Our simulations allowed us to observe the color-charge separation in
SU(3) fermionic systems in a generic non-commensurate case. %
We have obtained the charge and color velocities as a function of the
interaction from the real-time simulations both for the attractive and
repulsive case. %
Once we take into account finite size effects by renormalizing the
non-interacting velocities, our results at weak coupling are in
good agreement with bosonization calculations. %


The authors acknowledge discussions with Philippe Lecheminant and Stephan Rachel.
RAM is supported in part by Spanish Government grant
No.~FIS2009-07277, RT by a Feodor Lynen Fellowship of the Humboldt Foundation. 
We acknowledge the support by the Center  
for Functional Nanostructures (CFN), project B2.10. %

\bibliographystyle{prsty.bst} 

\end{document}